\documentclass[11pt]{article}
\usepackage[T1]{fontenc}
\usepackage[utf8]{inputenc}
\usepackage{ae}
\usepackage{aecompl}
\usepackage{lmodern}
\usepackage{amsmath}
\usepackage{amssymb}
\usepackage{graphicx}
\title{On the classification of consistent boundary conditions for $ \mathit{f}(\mathit{R})$-Gravity}
\author{H. Khodabakhshi $^{1}$, F. Shojai $^{1,2}$ and A. Shirzad $^{3,4}$ \\
\small$^1$Department of Physics, University of Tehran, Tehran, Iran\\
\small$^2$Foundations of Physics Group, School of Physics, \\ \small Institute for Research in Fundamental Sciences (IPM), \small Tehran, Iran\\ \small$^3$Department of Physics, Isfahan University of Technology, \small Isfahan, Iran\\ \small$^4$School of Particles and Accelerators, \\ \small Institute for Research in Fundamental Sciences (IPM),
\small Tehran, Iran\\
}
\date{}
\begin{document}
\makeatletter
\newcommand*{\rom}[1]{\expandafter\@slowromancap\romannumeral #1@}
\maketitle
%___________________________________________________________________________________________________
%
\begin{abstract}
Using a completely covariant approach, we discuss the role of boundary conditions (BCs) and the corresponding Gibbons--Hawking--York (GHY) terms in $ \mathit{f}(\mathit{R}) $-gravity  in arbitrary dimensions. We show that $ f(\mathit{R}) $-gravity, as a higher derivative theory, is not described by a degenerate Lagrangian, in its original form. Hence, without introducing additional variables, one can not obtain consistent BCs, even by adding the GHY terms (except for $f(\mathit{R})=R$).
 
However, following the Ostrogradsky approach, we can introduce a scalar field in the framework of Brans-Dicke formalism to the system to have consistent BCs by considering appropriate GHY terms. In addition to the Dirichlet BC, the GHY terms for both Neumann and two types of mixed BCs are derived. We show  the remarkable result that the $f(\mathit{R})$-gravity is itself compatible with one type of mixed BCs, in $D$ dimension, i.e. it doesn't require  any GHY term. For each BC, we rewrite the GHY term in terms of Arnowit-Deser-Misner (ADM) variables.
 
\end{abstract}
%___________________________________________________________________________________________________
%
\section{Introduction}
Since the theory of general relativity (GR) is a classical field theory of gravitation, the choice of BCs is of great importance. The role of surface integrals in GR has been investigated first in Dewitt and Dirac's papers \cite{dirac, dewitt} and then was covered deeply in the works of York, Regge and Teitelboim \cite{york, reggi}. In Ref. \cite{reggi}, the authors indicated the crucial role of the surface integral in order to have a well-defined functional space of the gravitational field. Three years later, trying to quantize GR in path integral formalism, Gibbons and Hawking \cite{GHY} showed that, a boundary term should be added to the Einstein-Hilbert (EH) action, in order to have a well-defined variational principle for an open manifold with Dirichlet BC, i.e. $ \delta g_{ab}|_{\textit{\tiny Boundary}}=0 $.  Such terms, added to the EH action, or the action of generalized theories of gravity \cite{Hint, fr, myers, madore, merino,krishna}, are called GHY terms.

The Lagrangian of GR as well as $ f(\mathit{R}) $-gravity contains second derivative of metric. Variational principle for these so called "Jerky mechanics" \cite{gold} is not well defined. In such actions, it is needed to apply Dirichlet and Neumann BCs simultaneously which may lead to destroy the Poisson structure of the system in a canonical treatment. However,  care is needed to define momentum and go to a well defined phase space via an ordinary Legendre transformation \cite{higher,higher1}. 

GR is described by  a degenerate Lagrangian, i.e. can be written as the sum of a quadratic part in the first derivatives of metric and a total derivative term. There are two approaches to deal with GR. The first one is the well-known ADM formalism which uses the Gauss-Codazzi equation to get rid of the second derivative terms of the Lagrangian \cite{GHY,Hint,fr,krishnan}. The second one, which is more covariant, uses the holographic relation to manifest the quadratic Lagrangian by subtracting a suitable  boundary term which can be removed by adding a GHY term \cite{padmanaban, padmanaban2006}.

For modified gravity models such as $f(\mathit{R})$-gravity one needs to use the so-called Ostrogradsky approach by introducing enough number of fields to the theory such that the whole Lagrangian of the system includes at most the first derivatives of the fields. In this way one is able to go through a canonical approach and at the same time introduce consistent BCs. For $f(\mathit{R})$-gravity without considering additional fields, one needs to consider the extrinsic curvature variation $\delta K_{ij}$, as well as  $\delta g_{ij}$ to vanish on the boundary, which is inconsistent since extrinsic curvature $ K_{ij}$, includes derivatives of the metric. However, by adding the famous GHY term $ - 2\int_{\partial \mathcal{M}}d^{D-1}y\epsilon\sqrt{h} f'(R)\delta \mathit{K} $ ($ h $ and $K$ are the trace of induced metric and the extrinsic curvature respectively, $ \epsilon=\pm 1 $ depending on the timelike or spacelike nature of the boundary $ \partial\mathcal{M} $ and $ f'(R)=\frac{\partial f}{\partial R} $) to the action, the BCs reduce to vanish $\delta R$ on the boundary simultaneously with the Dirichlet BC. But calculating $\delta R$ (see appendix B) shows that the main problem is not resolved since $R$ is not an independent field and its variation includes again variations of the derivatives of the metric.
  
In this paper we try in section 2 to investigate the more covariant approach for  $ f(\mathit{R}) $-gravity. For this reason, we use the equivalent scalar-tensor formulation of $ f(R) $-gravity and then using a suitable conformal transformation, we go to Einstein frame \cite{cap, so, st}. After imposing the holographic relation for Einsteinian curvature of space-time and changing back to the original variables \cite{fl}, we obtain somehow a holographic-like relation for $ f(\mathit{R}) $-gravity in which the bulk term is not quadratic. This shows that the $ f(\mathit{R}) $-gravity by itself is a non-degenerate Lagrangian and the ordinary approach is not suitable for it. 

Then in section 3, we try to change $ f(R) $ Lagrangian into a degenerate one by Ostrogradsky approach. To do so, we write the $ f(\mathit{R}) $-gravity in the Jordan frame of the  Brans-Dicke action \cite{cap, so, st}. Then by using the holographic relation for the curvature of space-time, we find that the action of the theory is degenerate, though there is not a well defined holographic relation. Hence, by adding appropriate GHY terms,  Dirichlet or other BCs can be achieved.  Writing the boundary terms of the action in terms of fields and momentum fields, in a foliation independent approach, enables us to introduce the consistent GHY term for Dirichlet, Neumann and two types of mixed BCs in arbitrary dimensions. For one type of mixed BC, the GHY term vanishes. This may be interpreted that the $f(\mathit{R})$-gravity is more consistent with this mixed type of BC in $D$ dimension.

In this paper the Latin indices are used to show the space-time coordinates and the Greek ones are used to denote the space coordinates. The calculations are done in arbitrary dimensions of space-time and the signature of metric is (-,+,+,+).

%___________________________________________________________________________________________________
\section{ Non-degeneracy of $ \mathit{f}(\mathit{R}) $-Gravity}
Similar to GR, $ f(\mathit{R}) $ Lagrangian includes second derivatives of the metric. The variational principle for this  type  of actions is not primarily well defined due to requirement of  applying simultaneously Dirichlet and Neumann BCs. There is also no room in this type of Lagrangian to define the momentum and establish Hamiltonian structure via Legendre transformation. In dealing with these Lagrangians, there are two possibilities: 1) The Lagrangian is degenerate i.e. it can be written as a quadratic Lagrangian which is equivalent to the original one plus a total derivative term which can be removed by adding a GHY term and imposing Dirichlet, Neumann or mixed BCs. 2) The Lagrangian is non-degenerate in which Ostrogradsky method or other equivalent methods should be used \cite{higher, higher1}.

In GR, Lagrangian is a degenerate Lagrangian, i.e. using the holographic relation, it can be written as
\begin{equation}
\sqrt{-g}R =\sqrt{-g}\mathcal{L}_{\textit{\tiny quad}}(\partial,\partial g)+\frac{1}{D/2-1}\partial_i\left(-g_{ab}\frac{\partial(\sqrt{-g}\mathcal{L}_{\textit{\tiny quad}})}{\partial(\partial_ig_{ab})}\right),
\label{5a}
\end{equation}
where 
\[ \mathcal{L}_{\textit{\tiny quad}}=\frac{1}{4}\mathcal{M}^{abcdef}\partial_ag_{bc}\partial_dg_{ef} \]
\begin{equation}
\mathcal{M}^{abcdef} \equiv g^{ad}\left(g^{bc}g^{ef}-g^{be}g^{cf}\right)+2g^{af}\left(g^{be}g^{cd}-g^{bc}g^{ed}\right)
\label{6a}
\end{equation}
As can be seen, it can be written as a quadratic Lagrangian plus a total derivative term \cite{padmanaban, padmanaban2006}. Unlike the GR Lagrangian, $ f(\mathit{R}) $-gravity given by 
\begin{equation}
\mathit{S}=\int d^Dx \sqrt{-g} f(\mathit{R}), 
\label{1}
\end{equation}
 seems to be non-degenerate. In order to obtain the holographic relation in $ f(\mathit{R}) $-gravity, first we try to write it in the GR form which we know how to work with it. To do so, we write the $ f(\mathit{R}) $ action using scalar-tensor theory as follows
\begin{equation}
\mathit{S}=\int d^{D}x\sqrt{-g}(\phi\mathit{R}-V(\phi)),
\label{p11}
\end{equation}
in which $ \phi=f'(\mathit{R}) $, $ V(\phi)=\mathit{R}(\phi)\phi-f(\mathit{R}(\phi)) $  and we have assumed that $ f''(R)\neq 0 $. This  is, in fact, the action of Brans-Dicke theory in the Jordan frame with parameter $ \omega=0 $ \cite{so, st, living, thomas}.
As is well-known, using the conformal transformation \cite{st}:
\begin{equation}
\tilde{g}_{ab}=\phi^{2/(D-2)}g_{ab}\ \ \ \ \ \ \ \ \ \ \ \ \ \ d\tilde{\phi}=\sqrt{\frac{2(D-1)}{(D-2)}}\frac{d\phi}{\phi}
\label{y11}
\end{equation}
the action (\ref{p11}) changes to Einstein gravity minimally coupled to a scalar field. Thus, in the so-called Einstein frame, the  separation of the Lagrangian into $ \text{bulk} $ and surface terms can be written as in Eq. (\ref{5a}). Then the obtained holographic relation can be restored into the Jordan frame by the inverse of transformation (\ref{y11}). To the end of this section, all quantities in the Einstein frame are denoted by $ \sim $. Noting 
\begin{equation}
\tilde{\mathit{R}}=\phi^{-2/(D-2)}\left(\mathit{R}-\frac{2(D-1)}{(D-2)}\frac{\square\phi}{\phi}+\frac{D-1}{D-2}\frac{\nabla_c\phi\nabla^c\phi}{\phi^2}\right)
\label{tilder}
\end{equation}
and  $ \sqrt{-\tilde{g}}=\phi^{D/(D-2)}\sqrt{-g} $, we can find 
\begin{equation}
\mathit{S}=\tilde{\mathit{S}}+\frac{2(D-1)}{D-2}\int_{\mathcal{M}}d^Dx\sqrt{-g}\square\phi,
\end{equation}
where 
\begin{equation}
\tilde{\mathit{S}}=\int_{\mathcal{M}}d^Dx\sqrt{-\tilde{g}}\left(\tilde{\mathit{R}}-\frac{1}{2}\tilde{\nabla}_a\tilde{\phi}\tilde{\nabla}^a\tilde{\phi}-U(\tilde{\phi})\right).
\label{circle}
\end{equation}
in which 
$ U(\tilde{\phi}(\phi))=\frac{V(\phi)}{\phi^{D/(D-2)}} $. 
Now we can separate the action of $ f(\mathit{R}) $-gravity into a quadratic $ \text{bulk} $ term and a surface term. To do this, let us recall that
\begin{equation}
\sqrt{-\tilde{g}}\tilde{\mathit{R}}=\sqrt{-\tilde{g}}\tilde{g}^{ab}\left(\tilde{\Gamma}^i_{ja}\tilde{\Gamma}^j_{ib}-\tilde{\Gamma}^i_{ab}\tilde{\Gamma}^j_{ij}\right)+\partial_c[\sqrt{-\tilde{g}}\tilde{V}^c],
 \label{ei}
\end{equation}
where $ \tilde{V}^c=\tilde{g}^{ik}\tilde{\Gamma}^c_{ik}-\tilde{g}^{ck}\tilde{\Gamma}^m_{mk} $ \cite{padmanaban}. Hence, the $ \text{bulk} $ term of (\ref{circle}) in the Einstein frame reads
\begin{equation}
\tilde{\mathcal{L}}_{\tiny\text{bulk}}=\tilde{g}^{ab}\left(\tilde{\Gamma}^i_{ja}\tilde{\Gamma}^j_{ib}-\tilde{\Gamma}^i_{ab}\tilde{\Gamma}^j_{ij}\right)-1/2\tilde{\nabla}_a\tilde{\phi}\tilde{\nabla}^a\tilde{\phi}-U(\tilde{\phi}).
\label{1111}
\end{equation}
The second term of (\ref{ei}) is denoted as $\tilde{\mathcal{L}}_{\tiny\text{Sur}}$ and leads to a surface term. 
Transforming back to the Jordan frame via Eq. (\ref{y11}), we obtain:
\begin{align}
 \sqrt{-\tilde{g}}\tilde{g}^{ab}\left(\tilde{\Gamma}^i_{ja}\tilde{\Gamma}^j_{ib}-\tilde{\Gamma}^i_{ab}\tilde{\Gamma}^j_{ij}\right)&=\phi\sqrt{-g}g^{ab}\left(\Gamma^i_{ja}\Gamma^j_{ib}-\Gamma^i_{ab}\Gamma^j_{ij}\right)\nonumber \\
 &+\phi\sqrt{-g}\left(\Gamma^i_{ij}\partial^i\ln\phi-g^{ab}\Gamma^i_{ab}\partial_i\ln\phi\right) \nonumber \\
&+\frac{D-1}{D-2}\phi\sqrt{-g}\partial_i\ln\phi\partial^i\ln\phi
\label{ho}
\end{align}  
and
\begin{equation}
\sqrt{-\tilde{g}}\tilde{V}^c=\phi\sqrt{-g}\left(g^{ik}\Gamma^c_{ik}-g^{ck}\Gamma^m_{km}\right)-2\frac{D-1}{D-2}\phi\sqrt{-g}\partial^c\ln\phi .
\label{yu}
\end{equation}
The above relations finally yield
\begin{equation}
\sqrt{-g}\mathcal{L}=\sqrt{-g}\mathcal{L}_{\tiny\text{bulk}}+\mathcal{L}_{\tiny\text{sur}}, 
\label{sep}
\end{equation}
where
\begin{align}
\mathcal{L}_{\tiny\text{bulk}}&=\phi g^{ab}\left(\Gamma^i_{ja}\Gamma^j_{ib}-\Gamma^i_{ab}\Gamma^j_{ij}\right)\nonumber \\
&+\phi\left(\Gamma^i_{ij}\partial^i\ln\phi-g^{ab}\Gamma^i_{ab}\partial_i\ln\phi\right)-V(\phi)
\label{s1s}
\end{align}
and
\begin{equation}
\mathcal{L}_{\tiny\text{sur}}=\partial_c(\phi\sqrt{-g}V^c)
\label{s2s}
\end{equation}
in which $ V^c=g^{ik}\Gamma^c_{ik}-g^{ck}\Gamma^i_{ik} $ \cite{fl, saltas}.

Now let us find the holographic relation for $ \mathit{f}(\mathit{R}) $-gravity similar to Eq. (\ref{5a}) for GR. It is clear that the holographic relation is satisfied in
the Einstein frame due to the minimal coupling of the scalar field to gravity. In order to write the holographic relation in the Jordan frame, let us start from Eqs. (\ref{s1s}) and (\ref{s2s}). A simple calculation shows
\begin{equation}
g_{ke}\frac{\partial(\sqrt{-g}\mathcal{L}_{\tiny\text{bulk}})}{\partial(\partial_cg_{ke})}=-\frac{D-2}{2}\phi\sqrt{-g}V^c+(D-1)\phi\sqrt{-g}\partial^c\ln\phi
\label{k}
\end{equation}
and
\begin{equation}
\mathcal{L}_{\tiny\text{sur}}=\partial_c(\phi\sqrt{-g}V^c)=-\frac{2}{D-2}\partial_c\left(g_{ke}\frac{\partial(\sqrt{-g}\mathcal{L}_{\tiny\text{bulk}})}{\partial(\partial_cg_{ke})}-(D-1)\sqrt{-g}\partial^c\phi\right),
\label{kh}
\end{equation}
Inserting (\ref{k}) and (\ref{kh}) into (\ref{sep}) and using $ \phi=f'(\mathit{R}) $, finally  one obtains
\begin{equation}
  \sqrt{-g}\mathcal{L}=\sqrt{-g}\mathcal{L}_{\tiny\text{bulk}}-\frac{2}{D-2}\partial_c\left(g_{ab}\frac{\partial(\sqrt{-g}\mathcal{L}_{\tiny\text{bulk}})}{\partial(\partial_cg_{ab})}-(D-1)\sqrt{-g}\partial^cf'(\mathit{R})\right)
\label{ssscircle}
\end{equation}
where
\begin{align}
 \mathcal{L}_{\tiny\text{bulk}}&=f'(\mathit{R})g^{ab}\left(\Gamma^i_{ja}\Gamma^j_{ib}-\Gamma^i_{ab}\Gamma^j_{ij}\right)\nonumber \\
 &+f'(\mathit{R})\left(\Gamma^j_{ij}\partial^i\ln f'(\mathit{R})-g^{ab}\Gamma^i_{ab}\partial_i\ln f'(\mathit{R})\right)\nonumber \\
&-(\mathit{R}f'(\mathit{R})-f(\mathit{R})),
\label{5656}
\end{align}
Considering relation (\ref{ssscircle}),  we see that the surface part of the Lagrangian is not determined completely by its bulk part. Therefore, we called it “holographic-like” relation. This is in contrast to EH Lagrangian, or more generally Lanczos-Lovelock Lagrangians \cite{fl}. 
Furthermore, the bulk Lagrangian in $ f(\mathit{R}) $-gravity is not  necessarily a quadratic Lagrangian and contains an arbitrary function of the second order derivatives of metric. Hence, the $ f(\mathit{R}) $ Lagrangian is not a degenerate Lagrangian.
%____________________________________________________________________________________________
\section{Ostrogradsky approach to  $f(\mathit{R})$-Gravity}
 As it was mentioned in the previous section, $ f(\mathit{R}) $ Lagrangian is not degenerate. Varying the action (\ref{1}) and  integrating by part, without implying any BC we have \footnote{ For a detailed calculations see Appendix A. }
\begin{align}
\delta \int_{\mathcal{M}} d^Dx\sqrt{-g}f(\mathit{R})&=\int_{\mathcal{M}} d^Dx\sqrt{-g} L^{ab}\delta g_{ab}\nonumber\\
&+\int_{\partial \mathcal{M}}d^{D-1}y \sqrt{h}\left\{-\frac{\mathit{f}^{\prime}\Pi_{ij}}{\sqrt{h}}+\epsilon\nabla_af'(h^a_jn_i-n^ah_{ij}) \right\}\delta h^{ij}\nonumber \\
&+\int_{\partial \mathcal{M}}d^{D-1}y\epsilon f'\delta(2\mathit{K}\sqrt{h})
\label{first}
\end{align}
where
\begin{equation}
L_{ab}\equiv-\frac{1}{2}\mathit{f}g_{ab}+\mathit{f}^{\prime}\mathit{R}_{ab}-\nabla_a\nabla_b\mathit{f}^{\prime}+g_{ab}\square \mathit{f}^{\prime}=0
\label{3}
\end{equation}
is the equation of motion. $\Pi_{ij}=\epsilon\sqrt{h}(\mathit{K}_{ij}-\mathit{K}h_{ij})$ is the momentum conjugate to the $ h^{ij} $ in GR and $ n_i $ is the normal vector of the boundary. As can be seen, to obtain the equations of motion, imposing the Dirichlet BC which leads to $ \delta h^{ij}|_{\textit{\tiny Boundary}}=0 $, we can get rid of the first surface integral in the above equation. To remove the second surface integral, there are two possibilities: 1) substituting $ \delta K_{ij}|_{\textit{\tiny Boundary}}=0 $, 2) adding the usual GHY boundary term to the action as follows 
\begin{equation}
\mathit{S}_t=\mathit{S}+\mathit{S}_{GHY}=\int_{\mathcal{M}} d^Dx \sqrt{-g} f(\mathit{R})-2\int_{\partial \mathcal{M}}d^{D-1}y\epsilon \sqrt{h} f' \mathit{K} .
\end{equation}
The first choice implies simultaneously vanishing of the metric and its derivatives on the boundary which is inconsistent. To investigate the second choice let us vary the above action 
\begin{align}
\delta \mathit{S}_t&=\int_{\mathcal{M}} d^Dx\sqrt{-g} L_{ab}\delta g^{ab}+\int_{\partial \mathcal{M}}d^{D-1}y\epsilon \sqrt{h}\left\{\frac{\mathit{f}^{\prime}\Pi_{ij}}{\sqrt{h}}+\nabla_af'(h^a_jn_i-n^ah_{ij}) \right\}\delta h^{ij}\nonumber \\
&-4\int_{\partial \mathcal{M}} d^{D-1}y\sqrt{h}\mathit{K} f' n_i \delta n^i-2\int_{\partial \mathcal{M}}d^{D-1}y\epsilon\sqrt{h}\mathit{K} f''\delta R.
\label{va}
\end{align}
Hence, to get the equations of motion, we need to impose $\delta R|_{\textit{\tiny Boundary}}=0 $, in addition to $ \delta h^{ij}|_{\textit{\tiny Boundary}}=\delta n^{i}|_{\textit{\tiny Boundary}}=0 $ which is Dirichlet BC because of $g_{ab} = h_{ab} +\epsilon  n_ a n_ b$. It should be noted that  although the normal to the boundary has a unit norm, this doesn't imply that the third term of (\ref{va}) is zero. A simple calculation shows that $ \delta n^b=\frac{1}{2} \epsilon n^bn_in_j\delta g^{ij}+h^b_i n_j \delta g^{ij} $, thus $ n_b\delta n^b=\frac{1}{2}n_in_j\delta g^{ij} $.   In appendix B we have shown that $\delta R$ is a combination of variations $ \delta h^{ij} , \delta K_{ij}, \delta n^i, \nabla_i \delta K $ and $ \delta (\nabla_a \nabla_i n^a ) $. Now we can ask if $ \delta R|_{\textit{\tiny Boundary}}=0 $ is compatible with the Dirichlet BC?
   
To answer this question we need to define, in a  consistent way, the momenta conjugate to the field variables in order to distinguish the Dirichlet and Neumann BCs where the momentum fields vanish on the boundary. However, this can be done only for degenerate theories, where the bulk term  contains at most the first order derivatives of the fields. Noting Ostrogradsky approach, we should change the $ f(\mathit{R}) $  Lagrangian into a degenerate one
    as much as possible. To do so, using scalar-tensor formulation, by introducing an scalar field $ \phi $, we write $ f(\mathit{R}) $ as  Lagrangian (\ref{p11}) which is not far form GR  that is degenerate.  Now substituting the holographic relation (\ref{5a}) in action (\ref{p11}), we have
\begin{equation}
\mathit{S}=\int_{\mathcal{M}}d^Dx \sqrt{-g}\left( \phi \mathcal{L}_{\textit{\tiny quad}}-V(\phi)\right)+\frac{1}{D/2-1} \int_{\mathcal{M}}d^Dx \phi \partial_i \left(-g_{ab}\frac{\partial(\sqrt{-g}\mathcal{L}_{\textit{\tiny quad}})}{\partial(\partial_{i} g_{ab})}\right)
\label{4747}
\end{equation}
The first integral contains only the metric, the field $ \phi $ and the first order derivatives of the metric. Integrating by parts, this is also the case for the second integral, and thus the above Lagrangian is degenerate. To see this, let us rewrite Eq. (\ref{4747}) as follows
\begin{align}
\mathit{S}&=\int_{\mathcal{M}}d^Dx \sqrt{-g}\left( \phi \mathcal{L}_{\textit{\tiny quad}}-V(\phi)\right)
+\frac{1}{D/2-1}\int_{\mathcal{M}}d^Dx \partial_i \phi g_{ab} \mathit{M}^{iab}\nonumber\\
&-\frac{1}{D/2-1}\int_{t}d^{D-1}y \phi g_{ab} \mathit{P}^{ab}.
\label{4848}
\end{align}
where
 $$ \mathit{M}^{iab} \equiv \frac{\partial(\sqrt{-g}\mathcal{L}_{\textit{\tiny quad}})}{\partial(\partial_{i} g_{ab})} = \frac{\sqrt{-g}}{2} \mathcal{M}^{iabpqr} \partial_p g_{qr} $$
and $ \mathcal{M}^{iabpqr} $ 
 is defined in Eq. (\ref{6a}). Note that $ \mathit{P}^{ab}\equiv \partial(\sqrt{-g}\mathcal{L}_{\textit{\tiny quad}})/\partial(\partial_{0} g_{ab})$ is the canonical momentum of $ g_{ab} $ in GR.
Hereafter we have also assumed that $ \partial \mathcal{M} $ contains two spacelike  $(D-1)$-dimensional surfaces at $ t=\text{constant} $ and one timelike surface on which the integral vanishes at large spatial distances. 
 Now we are able to define the canonical momenta of $ \phi $ and $ g_{ab} $ as follows 
 \begin{equation}
 \mathit{\bar{P}}^{ab} \equiv \frac{\delta\mathit{S}}{\delta (\partial_0 g_{ab})}=\phi \mathit{P}^{ab}+\frac{(D-1)\sqrt{-g}}{D-2}(g^{i0}g^{ab}-2g^{ib}g^{0a})\partial_i\phi
 \label{q0}
 \end{equation}
 and
\begin{equation}
\mathit{\bar{P}}_\phi \equiv \frac{\delta\mathit{S}}{\delta (\partial_0 \phi)}=\frac{1}{D/2-1}\mathit{P}
\label{q1}
\end{equation}
where $ \mathit{P} = g_{ab}\mathit{P}^{ab} $ and we have used the following relation
\begin{equation}
\partial_i \phi g_{ab} \mathcal{M}^{iab0qr} =(D-1)\{ g^{i0}g^{qr}-2g^{ir} g^{0q} \} \partial_i \phi.
\label{q2}
\end{equation}
Considering the action (\ref{4848}), one can see that, regardless of the surface integral which is a GHY term, 
the Lagrangian contains fields and their first order derivatives. Therefore, we can be sure that the variational principle for this action is compatible with the Dirichlet BC. Before investigating 
in details the compatibility of the model, let us show explicitly the structure of the added GHY term in the ADM formalism. Consider the following relation
\begin{align}
\mathit g_{ab}{P}^{ab}&=\frac{\sqrt{-g}}{4}g_{ab}g_{de,f}\left[\mathcal{M}^{ab0def}+\mathcal{M}^{defab0}\right]\nonumber\\
 &=\sqrt{-g}\frac{D-2}{2}g_{de,f}\left(g^{de}g^{0f}-g^{df}g^{0e}\right)=\frac{D-2}{2}\frac{1}{\sqrt{-g}}\partial_a(gg^{0a})\nonumber\\
&=\frac{D-2}{2}\sqrt{-g}\left(\frac{-1}{N^2}\partial_a(N^2g^{0a})-g^{0a}\partial_a\ln h\right)\nonumber \\
&=\frac{D-2}{2}\sqrt{-g}\left[-2\mathit{K}n^0+\frac{\partial_{\alpha}\mathit{N}^{\alpha}}{\mathit{N}^2}\right]
\label{f}
\end{align}
where $n^a=N^{-1}(1,-N^\alpha)$ and the lapse and shift functions are denoted by $ N $ and $ N^{\alpha}$. In the last equality we have used the following two identities
\begin{equation}
2\mathit{K}n^o=\frac{2}{N^2\sqrt{h}}\partial_a(N^2\sqrt{h}g^{0a})=\frac{2}{N^2}\partial_a(N^2g^{0a})+2g^{0a}\partial_a\ln\sqrt{h}
\end{equation} 
\begin{equation}
\frac{1}{N^2}\partial_a(N^2g^{0a})=\frac{\partial_{\alpha}N^{\alpha}}{N^2} 
\end{equation} 
Substituting the expression (\ref{f}) into (\ref{4848}) gives
\begin{align}
\mathit{S}&=\int_{\mathcal{M}}d^Dx \sqrt{-g}\left( \phi \mathcal{L}_{\textit{\tiny quad}}-V(\phi)\right)+\frac{1}{D/2-1}\int_{\mathcal{M}}d^Dx \partial_i \phi g_{ab} \mathit{M}^{iab}\nonumber\\
&-2\int_{t}d^{D-1}y\sqrt{h} \phi \mathit{K}+\int_{t}d^{D-1}y\sqrt{h}\phi \frac{\partial_{\alpha}N^{\alpha}}{N}
\label{494949}
\end{align}
where the first surface integral in (\ref{494949}), is the same as GHY term of Refs. \cite {Hint, fr}. However, the second surface term is often lost in the literatures. We will come back to this point in the next subsection.

Now let us consider the  variations of the action (\ref{4848}). First, we rewrite it in terms of the momenta given in Eqs. (\ref{q0}) and (\ref{q1}). By adding and subtracting the following surface integral
\begin{equation}
\frac{2(D-1)}{(D-2)^2}\int_t d^{D-1}y \sqrt{-g} g_{ab} (g^{i0}g^{ab}-2g^{ib}g^{0a})\partial_i\phi=\frac{2(D-1)}{D-2}\int_t d^{D-1}y \sqrt{-g} \partial^0\phi
\end{equation}
to the action (\ref{4848}), we get
\begin{align}
\mathit{S}&=\int_{\mathcal{M}}d^Dx \sqrt{-g}\left( \phi \mathcal{L}_{\textit{\tiny quad}}-V(\phi)\right)
+\frac{2}{D-2}\int_{\mathcal{M}}d^Dx \partial_i \phi g_{ab} \mathit{M}^{iab}\nonumber \\
&-\frac{2}{D-2}\int_{t}d^{D-1}y g_{ab} \mathit{\tilde{P}}^{ab}+\frac{2(D-1)}{D-2}\int_t d^{D-1}y \sqrt{-g} \partial^0\phi
\label{48480}
\end{align}
Varying this action with respect to $ \phi $ and $ g_{ab} $ and using Eq. (\ref{q2}), after a little algebra, we obtain
\begin{equation}
\delta \mathit{S}= \delta_{\phi}\mathit{S}+\delta_{g}\mathit{S}
\label{9a}
\end{equation}
where
\begin{equation}
\delta_{\phi}\mathit{S}=\int_{\mathcal{M}}d^Dx \left\{\sqrt{-g}\left( \mathcal{L}_{\textit{\tiny quad}}-\partial_\phi V(\phi)\right) +\frac{1}{D/2-1} \partial_i \left(-g_{ab}\frac{\partial(\sqrt{-g}\mathcal{L}_{\textit{\tiny quad}})}{\partial(\partial_{i} g_{ab})}\right)\right\}\delta \phi
\label{8a}
\end{equation} 
and
\begin{align}
\delta_{g}\mathit{S}&=\int_{\mathcal{M}}d^Dx \mathit{L}^{ab} \delta g_{ab} + \frac{D-4}{D-2}\int_{t}d^{D-1}y \mathit{\bar{P}}^{ab} \delta g_{ab}\nonumber \\
&+\frac{D-1}{D-2}\int_{t}d^{D-1}y \sqrt{-g} \partial^0 \phi g^{ab} \delta g_{ab}-\frac{2}{D-2}\int_{t}d^{D-1}y g_{ab} \delta \mathit{\bar{P}}^{ab}\nonumber\\
&+\frac{2}{D-2}\int_{t}d^{D-1}y g_{ab}\mathit{P}^{ab} \delta \phi+\frac{2(D-1)}{D-2}\int_{t}d^{D-1}y \sqrt{-g} \delta(\partial^0 \phi)
\label{5353} 
\end{align}
in which
\begin{align}
 \mathit{L}^{ab}&=\phi \frac{ \partial(\sqrt{-g}\mathcal{L}_{\textit{\tiny quad}})}{\partial g_{ab}} - \partial _i\left(\phi \mathit{M}^{iab}\right)-\frac{1}{2}\sqrt{-g} g^{ab} V(\phi)+\frac{2}{D-2} \partial_i \phi \mathit{M}^{iab} \nonumber \\
 &+\frac{2}{D-2}\partial_i \phi g_{kl}\mathit{H}^{iabkl}-\frac{1}{D-2} \partial_p (\sqrt{-g}\partial_i \phi g_{qr} \mathcal{M}^{iqrpab})
 \end{align}
and $\mathit{H}^{iabkl} \equiv \partial\mathit{M}^{iab}/\partial g_{kl} $.
Substituting (\ref{8a}) and (\ref{5353}) in (\ref{9a}) gives
\begin{align}
\delta \mathit{S}&= \int_{\mathcal{M}}d^Dx \bigg\{ \sqrt{-g}\left( \mathcal{L}_{\textit{\tiny quad}}-\partial_\phi V(\phi)\right) +\frac{1}{D/2-1} \partial_i \left(-g_{ab}\frac{\partial(\sqrt{-g}\mathcal{L}_{\textit{\tiny quad}})}{\partial(\partial_{i} g_{ab})}\right)\bigg \}\delta \phi \nonumber \\
&+\int_{\mathcal{M}}d^Dx \mathit{L}^{ab} \delta g_{ab} + \frac{D-4}{D-2}\int_{t}d^{D-1}y \mathit{\bar{P}}^{ab} \delta g_{ab}-\frac{2}{D-2}\int_{t}d^{D-1}y g_{ab} \delta \mathit{\bar{P}}^{ab}\nonumber \\
 &+\frac{D-1}{D-2}\int_{t}d^{D-1}y \sqrt{-g} \partial^0 \phi g^{ab} \delta g_{ab}
 +\int_{t}d^{D-1}y \bar{\mathit{P}}_\phi \delta \phi \nonumber \\
 &+\frac{2(D-1)}{D-2}\int_{t}d^{D-1}y \sqrt{-g} \delta(\partial^0 \phi) 
\label{vari} 
 \end{align}
As expected, without the GHY term, the  undesirable BCs: $ \delta g_{ab}|_{\textit{\tiny Boundary}}=\delta \mathit{\bar{P}}_{ab}|_{\textit{\tiny Boundary}}= \delta \phi|_{\textit{\tiny Boundary}}=\delta (\partial^0 \phi)|_{\textit{\tiny Boundary}}=0$   should be assigned. In order to find  the appropriate GHY term, let us discuss three different types of BCs leading to a consistent stationary action principle for $ \mathit{f}(\mathit{R})$-gravity.
%____________________________________________________________________________________________________
\subsection{Dirichlet BC}
Considering the surface terms in Eq. (39), in order to impose the Dirichlet BC $ \delta g_{ab}|_{\textit{\tiny Boundary}}=\delta \phi|_{\textit{\tiny Boundary}}=0 $, we need to modify the action (\ref{48480})  by adding the following GHY term
\begin{equation}
\mathit{S}_{\textit{\tiny D}}=\mathit{S}+\mathit{S}_{\textit{\tiny D}} ^{\textit{\tiny GHY}}=\mathit{S}+\frac{2}{D-2}\int_{t}d^{D-1}y g_{ab} \mathit{\bar{P}}^{ab}-\frac{2(D-1)}{D-2}\int_t d^{D-1}y \sqrt{-g} \partial^0 \phi
\end{equation}
To see that the above action is compatible with the Dirichlet BC, let us vary it as follows
\begin{align}
\delta \mathit{S}_{\textit{\tiny D}}&= \int_{\mathcal{M}}d^Dx \left\{ \sqrt{-g}\left( \mathcal{L}_{\textit{\tiny quad}}-\partial_\phi V(\phi)\right) +\frac{1}{D/2-1} \partial_i \left(-g_{ab}\frac{\partial(\sqrt{-g}\mathcal{L}_{\textit{\tiny quad}})}{\partial(\partial_{i} g_{ab})}\right)\right\}\delta \phi \nonumber \\
&+\int_{\mathcal{M}}d^Dx \mathit{L}^{ab} \delta g_{ab} + \int_{t}d^{D-1}y \mathit{\bar{P}}^{ab} \delta g_{ab}+\int_{t}d^{D-1}y \bar{P}_{\phi}\delta \phi
\label{com}
\end{align}
which gives the equation of motion subjected the Dirichlet BC. Note that $ \phi=f'(R) $ gives $ \delta \phi|_{\textit{\tiny Boundary}}=f''(R)\delta R|_{\textit{\tiny Boundary}}=0 $. Now we can surely say that $ \delta R|_{\textit{\tiny Boundary}}=0 $ is compatible with the Dirichlet BC and is in fact part of it. This is a clear covariant verification of the result pointed in Ref. \cite{Hint} in the framework of the ADM foliation. To be more concrete, we can determine the GHY term $S_{\textit{\tiny D}} ^{\textit{\tiny GHY}}$ in terms of ADM variables. 
Using (\ref{f}) and substituting $ \phi= f'(R) $, we have
\begin{equation}
\mathit{S}_{\textit{\tiny D}} ^{\textit{\tiny GHY}}=-2\int_{t}d^{D-1}y\sqrt{h} f'(R) \mathit{K}+\int_{t}d^{D-1}y\sqrt{h} f'(R) \frac{\partial_{\alpha}N^{\alpha}}{N}
\label{p0}
\end{equation}
which are the same terms present in Eq. (\ref{494949}). Since in ADM formalism, the Dirichlet BC means $ \delta h^{ab}|_{\textit{\tiny Boundary}}= \delta N^\mu|_{\textit{\tiny Boundary}}=\delta N|_{\textit{\tiny Boundary}}=0 $, the last term of the above equation can be neglected and  the first term suffices.  However, note that this is correct only for the Dirichlet BC. 

To complete our discussion, we can set $ \phi=1 $ and $ V(\phi)=0 $ in Eq. (\ref{vari}) to find the following result for the case of GR 
\begin{equation}
\delta S_{\textit{\tiny (EH)}}= \int_{\mathcal{M}}d^Dx \mathit{\bar{L}}^{ab} \delta g_{ab} + \frac{D-4}{D-2}\int_{t}d^{D-1}y \mathit{P}^{ab} \delta g_{ab}-\frac{2}{D-2}\int_{t}d^{D-1}y g_{ab} \delta \mathit{P}^{ab}
\end{equation}
where
\[ \mathit{\bar{L}}^{ab}= \frac{ \partial(\sqrt{-g}\mathcal{L}_{\textit{\tiny quad}})}{\partial g_{ab}} - \partial _i \mathit{M}^{iab}\]
Imposing the Dirichlet BC: $ \delta g_{ab}|_{\textit{\tiny Boundary}}=0 $,  the action should be modified by the following GHY term 
to get the equations of motion,   %
\begin{equation}
\mathit{S}_{\textit{\tiny D(EH)}}=\mathit{S}_{\textit{\tiny (EH)}}+\mathit{S}_{\textit{\tiny D(EH)}} ^{\textit{\tiny GHY}}=\mathit{S}_{\textit{\tiny (EH)}}+\frac{2}{D-2}\int_{t}d^{D-1}y g_{ab} \mathit{P}^{ab}
\label{h01}
\end{equation}
 Moreover, using  Eq. (\ref{f}), we can rewrite $\mathit{S}_{\textit{\tiny D(EH)}} ^{\textit{\tiny GHY}}$ in the familiar form
\begin{equation}
\mathit{S}_{\textit{\tiny D(EH)}} ^{\textit{\tiny GHY}}=-2\int_{t}d^{D-1}y \sqrt{h} \mathit{K} + \int_{t}d^{D-1}y \sqrt{h} \frac{\partial_{\alpha} N^{\alpha}}{N}
\label{h1}
\end{equation}
where for the Dirichlet BC, the second term can be neglected \cite{Poisson}.
%____________________________________________________________________________________________________
\subsection{Neumann BC}
In order to obtain the GHY term related to the Neumann BC: $ \delta \bar{P}^{ab}|_{\textit{\tiny Boundary}}=\delta \bar{P}_\phi|_{\textit{\tiny Boundary}}=0 $, let us write (\ref{vari}) in a different form. From (\ref{q0}) and (\ref{q1}), we find that
\begin{equation}
\bar{P}^{ab}\delta g_{ab}=-g_{ab}\delta\bar{P}^{ab}+\frac{D-2}{2}\bar{P}_\phi \delta \phi +\frac{D-2}{2} \phi \delta\bar{P}_\phi + (D-1) \delta (\sqrt{-g}\partial^0 \phi)
\label{mn0} 
\end{equation}
Inserting this into (\ref{vari}) gives
\begin{align}
\delta \mathit{S}&= \int_{\mathcal{M}}d^Dx \bigg\{ \sqrt{-g}\left( \mathcal{L}_{\textit{\tiny quad}}-\partial_\phi V(\phi)\right) +\frac{1}{D/2-1} \partial_i \left(-g_{ab}\frac{\partial(\sqrt{-g}\mathcal{L}_{\textit{\tiny quad}})}{\partial(\partial_{i} g_{ab})}\right)\bigg \}\delta \phi \nonumber \\
&+\int_{\mathcal{M}}d^Dx \mathit{L}^{ab} \delta g_{ab} - \int_{t}d^{D-1}y g_{ab}  \delta \mathit{\bar{P}}^{ab} + \frac{D-4}{2}\int_{t}d^{D-1}y \phi \delta \mathit{\bar{P}}_\phi \nonumber \\
 &+\frac{D-2}{2}\int_{t}d^{D-1}y \mathit{\bar{P}}_\phi \delta \phi+ \frac{D-1}{2}\int_{t}d^{D-1}y \sqrt{-g} \partial^0 \phi g^{ab} \delta g_{ab}\label{varii}
  \nonumber \\
 &+{(D-1)}\int_{t}d^{D-1}y \sqrt{-g} \delta(\partial^0 \phi) 
 \end{align}
This shows that the action (\ref{48480}) is consistent with the Neumann BC if we propose the following GHY term
\begin{equation}
\mathit{S}_{\textit{\tiny N}}=\mathit{S}+\mathit{S}_{\textit{\tiny N}} ^{\textit{\tiny GHY}}=\mathit{S}-\frac{D-2}{2}\int_{t}d^{D-1}y \mathit{\bar{P}}_\phi \phi -(D-1)\int_t d^{D-1}y \sqrt{-g} \partial^0 \phi
\label{N}
\end{equation}
Variation of (\ref{N}) yields
\begin{align}
\delta \mathit{S}_{\textit{\tiny N}}&= \int_{\mathcal{M}}d^Dx \left\{ \sqrt{-g}\left( \mathcal{L}_{\textit{\tiny quad}}-\partial_\phi V(\phi)\right) +\frac{1}{D/2-1} \partial_i \left(-g_{ab}\frac{\partial(\sqrt{-g}\mathcal{L}_{\textit{\tiny quad}})}{\partial(\partial_{i} g_{ab})}\right)\right\}\delta \phi \nonumber \\
&+\int_{\mathcal{M}}d^Dx \mathit{L}^{ab} \delta g_{ab} - \int_{t}d^{D-1}y g_{ab} \delta\mathit{\bar{P}}^{ab} - \int_{t}d^{D-1}y \phi \delta \bar{P}_{\phi}
\end{align}
which gives the equations of motion using Neumann BC.

Using (\ref{f}) and inserting $ \phi=f'(R) $, we can write the GHY term in (\ref{N}) in the ADM formalism as
\begin{align}
\mathit{S}_{\textit{\tiny N}} ^{\textit{\tiny GHY}}&=(D-2)\int_{t}d^{D-1}y \sqrt{h} f'(R) \mathit{K}-\frac{D-2}{2} \int_{t}d^{D-1}y \sqrt{h} f'(R) \frac{\partial_{\alpha} N^{\alpha}}{N}\nonumber\\
&-(D-1)\int_td^{D-1}y\sqrt{h}N\partial^0 f'(R)
\label{p1}
\end{align}
It should be noted that unlike the case of Dirichlet BC, the second term in the above action can not be neglected unless for the coordinate system in which $ N^{\alpha}=0 $. It is worth to compare (\ref{p1}) with the GHY term  (\ref{p0}) for Dirichlet BC. It is easily seen that
\begin{equation}
\mathit{S}_{\textit{\tiny N}} ^{\textit{\tiny GHY}}=-\frac{D-2}{2}\mathit{S}_{\textit{\tiny D}} ^{\textit{\tiny GHY}}-(D-1)\int_td^{D-1}y\sqrt{h}N\partial^0 f'(R)
\end{equation}

%____________________________________________________________________________________________________
\subsection{Mixed BC}
There are two types of mixed BCs for $f(R)$-gravity: $ \delta\bar{P}_{ab}|_{\textit{\tiny Boundary}}=\delta \phi|_{\textit{\tiny Boundary}}=0$ or $ \delta \bar{P}_\phi|_{\textit{\tiny Boundary}}=\delta g_{ab}|_{\textit{\tiny Boundary}}=0$. We begin with the first one. Using the variation of $f(R)$-gravity action, (\ref{vari}) or (\ref{varii}), the first type mixed BC would be consistent if we have added the following GHY term to the action
\begin{equation}
\mathit{S}^{\tiny}_{\textit{\tiny MI}}=\mathit{S}+\mathit{S}_{\textit{\tiny MI}} ^{\textit{\tiny GHY}}=\mathit{S}-\frac{D-4}{2}\int_{t}d^{D-1}y \phi \mathit{\bar{P}}_\phi -(D-1)\int_{t}d^{D-1}y \sqrt{-g} \partial^0 \phi
\label{z0}
\end{equation}
Varying the above action gives
\begin{align}
\delta \mathit{S}_{\textit{\tiny MI}}&=\int_{\mathcal{M}}d^Dx \left\{ \sqrt{-g}\left( \mathcal{L}_{\textit{\tiny quad}}-\partial_\phi V(\phi)\right) +\frac{1}{D/2-1} \partial_i \left(-g_{ab}\frac{\partial(\sqrt{-g}\mathcal{L}_{\textit{\tiny quad}})}{\partial(\partial_{i} g_{ab})}\right)\right\}\delta \phi \nonumber\\
&+\int_{\mathcal{M}}d^Dx \mathit{L}^{ab} \delta g_{ab} - \int_{t}d^{D-1}y g_{ab} \delta \mathit{\bar{P}}^{ab}+\int_{t}d^{D-1}y \bar{P}_{\phi} \delta \phi
\end{align}
As can be seen, the mixed BC: $ \delta \mathit{\bar{P}}_{ab}|_{\textit{\tiny Boundary}}=\delta \phi|_{\textit{\tiny Boundary}}=0 $  yields consistently the equations of motion.
Now using Eq. (\ref{f}) and $ \phi=f'(R) $, we can write the GHY term of Eq.  (\ref{z0}) in terms of ADM variables as
\begin{align}
\mathit{S}_{\textit{\tiny MI}} ^{\textit{\tiny GHY}}&=(D-4)\int_{t}d^{D-1}y \sqrt{h} f'(R) \mathit{K}-\frac{D-4}{2} \int_{t}d^{D-1}y \sqrt{h} f'(R) \frac{\partial_{\alpha} N^{\alpha}}{N}\nonumber\\
&-(D-1)\int_td^{D-1}y\sqrt{h}N\partial^0 f'(R)
\label{mix1}
\end{align}
 
It is also worth noting here to find the relation between Dirichlet and the above mixed GHY boundary terms in $ f(R) $-gravity. Comparing the above result with that obtained in Eq. (\ref{p0}), we see that 
\begin{equation}
\mathit{S}_{\textit{\tiny MI}} ^{\textit{\tiny GHY}}=-\frac{D-4}{2}\mathit{S}_{\textit{\tiny D}} ^{\textit{\tiny GHY}}-(D-1)\int_td^{D-1}y\sqrt{h}N\partial^0 f'(R)
\end{equation}
For GR, i.e. $ \phi=1 $ and $ V(\phi)=0 $,  it is interesting that the newly defined action (\ref{z0}) is consistent with the Neumann BC
\begin{equation}
\mathit{S}_{\textit{\tiny N(EH)}}=\mathit{S}_{\textit{\tiny (EH)}}+\mathit{S}_{\textit{\tiny N(EH)}} ^{\textit{\tiny GHY}}=\mathit{S_{\textit{\tiny (EH)}}}-\frac{D-4}{D-2}\int_{t}d^{D-1}y P
\label{h0}
\end{equation}
This shows that the pure Neumann BC may be used for GR in arbitrary dimensions.  Moreover in four dimensions, for GR with Neumann BC, there is no need to any GHY term in order to have a consistent theory.
%GR is a self-consistent theory of gravity with the Neumann BC  with no boundary terms. 
This point, explained here covariantly is also shown recently in \cite{krishnan} in ADM approach.
%This point that derived here covariantly, recently mentioned in the ADM approach in Ref. \cite{krishnan}. 
To clarify more, using Eq. (\ref{f}), it is easy to see that the above GHY term with respect to the ADM variables takes the form  
\begin{equation}
\mathit{S}_{\textit{\tiny N(EH)}} ^{\textit{\tiny GHY}}=(D-4)\int_{t}d^{D-1}y \sqrt{h} \mathit{K} -\frac{D-4}{2} \int_{t}d^{D-1}y \sqrt{h} \frac{\partial_{\alpha} N^{\alpha}}{N}.
\label{h2}
\end{equation}
where again the second term in the above action or in (\ref{mix1}), can be ignored only for the special choice of coordinate system mentioned in the previous subsection. Also, it can be easily seen that, in GR, in contrast to the Dirichlet case, the required GHY term, compatible with the Neumann BC, depends on the dimension of space-time and for $ D=4 $, the coefficient of GHY term vanishes, as expected. Another interesting feature is the relation between Dirichlet and  Neumann GHY term in GR. Comparing Eqs. (\ref{h1}) and (\ref{h2}), one finds
\begin{equation}
\mathit{S}_{\textit{\tiny N(EH)}}^{\textit{\tiny GHY}}=-\frac{D-4}{2}\mathit{S}_{\textit{\tiny D(EH)}}^{\textit{\tiny GHY}}
\end{equation}

Now let us look at the second type of mixed BC: $ \delta \bar{P}_\phi|_{\textit{\tiny Boundary}}=\delta g^{ab}|_{\textit{\tiny Boundary}}=0 $. In order to discuss the consistency of $ \mathit{f}(\mathit{R})$-gravity with this BC, first we use (\ref{mn0}) to substitute for $g_{ab}\delta\bar{P}^{ab} $ in (\ref{vari}). This leads to
\begin{align}
\delta \mathit{S}_{\textit{\tiny MII}}&=\int_{\mathcal{M}}d^Dx \left\{ \sqrt{-g}\left( \mathcal{L}_{\textit{\tiny quad}}-\partial_\phi V(\phi)\right) +\frac{1}{D/2-1} \partial_i \left(-g_{ab}\frac{\partial(\sqrt{-g}\mathcal{L}_{\textit{\tiny quad}})}{\partial(\partial_{i} g_{ab})}\right)\right\}\delta \phi \nonumber\\
&+\int_{\mathcal{M}}d^Dx \mathit{L}^{ab} \delta g_{ab} + \int_{t}d^{D-1}y  \mathit{\bar{P}}^{ab}\delta g_{ab} - \int_{t}d^{D-1}y \phi \delta \bar{P}_{\phi}
\label{mn3}
\end{align}
Clearly by applying the BC: $ \delta \bar{P}_{\alpha}|_{\textit{\tiny Boundary}}=\delta g_{ab}|_{\textit{\tiny Boundary}}=0 $, we can get the equations of motion without adding any GHY term to the above expression. This means that $ f(R) $-gravity with the above type of mixed BC is self-consistent with no need to any GHY term in $ D $ dimension. 
%This means that the $ f(R) $-gravity action is self-consistent with Neamann BC mentioned above in $ D $ dimension. 
 To clarify this point better, let us return to the relation (\ref{first}) in which the boundary terms are written in the ADM formalism. One can write these terms in term of the momenta conjugate to $ \phi $, $ h_{ij} $, $ N $ and $ N^\alpha $. These are derived in details in appendix C and are as follows
\[ \bar{\Pi}_N=\bar{\Pi}_{N^{\alpha}}=0 \]
\[ \bar{\Pi}_\phi=-2\epsilon \sqrt{h}\mathit{K} \]
 \begin{equation}
 \bar{\Pi}_{ij}=\epsilon \sqrt{h}\left\{\phi(\mathit{K}_{ij}-\mathit{K}h_{ij})+\frac{h_{ij}}{N}(\dot{\phi}-N^{\alpha}\partial_\alpha\phi)\right\}
 \end{equation}
Substituting this into (\ref{first}) and inserting $ \phi=f'(R) $, we obtain
\begin{align}
\delta \int_{\mathcal{M}} d^Dx\sqrt{-g}f(\mathit{R})&=\int_{\mathcal{M}} d^Dx\sqrt{-g} L^{ab}\delta g_{ab}\nonumber\\
&-\int_{\partial \mathcal{M}}d^{D-1}y \sqrt{h}\bar{\Pi}_{ij}\delta h^{ij}
 -\int_{\partial \mathcal{M}}d^{D-1}yf'\delta\bar{\Pi}_\phi \nonumber \\
&-\int_{\partial \mathcal{M}}d^{D-1}y\epsilon \sqrt{h} \nabla^i\phi\delta n_i
\label{2}
\end{align}
where $ h^{ij}\delta n_i =-n_i\delta h^{ij}$ is used. It can be seen that the above surface terms, which are written in the ADM formalism, are  completely in agreement with what we have derived by the covariant approach in (\ref{mn3}). Regarding the above relation and by applying the mixed BC: $ \delta \bar{\Pi}_\phi|_{\textit{\tiny Boundary}}=\delta n_i|_{\textit{\tiny Boundary}}=\delta h^{ij}|_{\textit{\tiny Boundary}}=0 $,  we can get the equations of motion in $ D $ dimension without any GHY term.
%
%_________________________________________________________________________________
\section{conclusion}
In this paper it is shown that unlike GR, the Lagrangian of $ f(\mathit{R}) $-gravity does not follow a holographic relation which is the feature of the Lanczos-Lovelock Lagrangian. Moreover, the Lagrangian of $ f(\mathit{R}) $-gravity can not be expressed as the sum of quadratic and total derivative terms. So $ f(\mathit{R}) $ Lagrangian is not degenerate. Following the Ostrogradsky approach, since $ f(\mathit{R}) $-gravity is a theory with higher order derivatives of metric, it carries a single additional degree of freedom, which is the scalar field of equivalent Brans-Dicke action. Introducing this  field, leads to a degenerate Lagrangian which is used to develop the problem of BC and the corresponding GHY terms in $ f(\mathit{R}) $-gravity \cite{Hint, fr}.
 
Here we have followed a foliation independent approach to find the GHY boundary terms in $ f(\mathit{R}) $-gravity, required to make the BC variation problem well-defined. We have shown that in addition to the Dirichlet BC, the Neumann BC and two types of the mixed BCs can be introduced for the $f(\mathit{R})$-gravity. The remarkable point which is \textbf{one} of the main results of this paper is about the mixed BCs. We have shown that one of the mixed BC: $ \delta\bar{P}_{ab}|_{\textit{\tiny Boundary}}=\delta \phi|_{\textit{\tiny Boundary}}=0$ is reduced to the Neumann BC in the case of GR.  This BC together the other mixed BC: $\delta \bar{P}_\phi|_{\textit{\tiny Boundary}}=\delta g^{ab}|_{\textit{\tiny Boundary}}=0$ are self-consistent BCs, i.e. these do not need to any GHY term to be consistent with the theory, the first one for GR and the second one for $f(\mathit{R})$-gravity, both in $D$ dimension. 
\appendix
\renewcommand{\theequation}{\thesection.\arabic{equation}}
\setcounter{equation}{0}
%______________________________________________________________________________________________________
\section{Variation of $ f(\mathit{R}) $-gravity action without BC}
The variation of the action of $ f(\mathit{R}) $-gravity gives
\begin{align}
\delta\mathit{S}_f&=\delta\int_{\mathcal{M}} d^Dx\sqrt{-g}f(\mathit{R})\nonumber\nonumber \\
& =\int_{\mathcal{M}} d^Dx\sqrt{-g}\left({-\frac{1}{2}fg_{ab}+f'\mathit{R}_{ab}}\right)\delta g^{ab}+\int \sqrt{-g}f'g^{ab}\delta \mathit{R}_{ab}
\label{a1}
\end{align}
The first integral includes some terms of the equations of motion. Using the contracted form of Palatini equation 
\begin{equation}
g^{ik}\delta\mathit{R}_{ik}=\nabla_a\left(g^{ik}\delta\Gamma^a_{ik}-g^{ia}\delta\Gamma^k_{ik}\right)=\nabla_a\nabla_b\left(-\delta g^{ab}+g^{ab}g_{ik}\delta g^{ik}\right),
\end{equation}
and integrating by part in the second term of (\ref{a1}), we would have
\begin{align}
\int_{\mathcal{M}} d^D x\sqrt{-g}f'g^{ab}\delta \mathit{R}_{ab}&=\int_{\mathcal{M}}d^{D}x\sqrt{-g}(\nabla_d\nabla_af')(-\delta g^{ad}+g^{ad}g_{ik}\delta g^{ik})\nonumber \\
&+\int_{\partial \mathcal{M}} d^{D-1}y\epsilon\sqrt{h}n_af'\nabla_d(-\delta g^{ad}+g^{ad}g_{ik}\delta g^{ik})\nonumber \\
&-\int_{\partial \mathcal{M}} d^{D-1}y\epsilon\sqrt{h}n_a(\nabla_df')(-\delta g^{ad}+g^{ad}g_{ik}\delta g^{ik})
\label{a2}
\end{align}
Inserting (\ref{a2}) in (\ref{a1}), we get
\begin{align}
\delta\mathit{S}_{\mathit{f}}&=\int_{\mathcal{M}} d^D x \sqrt{-g}\left[-\frac{1}{2}\mathit{f}g_{ab}+\mathit{f}^{\prime}\mathit{R}_{ab}-\nabla_a\nabla_b\mathit{f}^{\prime}+g_{ab}\square \mathit{f}^{\prime}\right]\delta g^{ab}\nonumber \nonumber \\
&+\int_{\partial \mathcal{M}} d^{D-1}y\sqrt{h}\epsilon n_a\bigg\{f'\left(\nabla_d(-\delta g^{ad}+g^{ad}g_{ik}\delta g^{ik}) \right)\nonumber \\
&-\left(\nabla_df'(-\delta g^{ad}+g^{ad}g_{ik}\delta g^{ik})\right)\bigg\}
\label{nmn}
\end{align}
Now we want to write the surface integral of (\ref{nmn}) in ADM foliation of space-time. The first term gives
\begin{align}
&\int_{\partial \mathcal{M}}d^{D-1}y\epsilon\sqrt{h}n_af'\nabla_d(-\delta g^{ad}+g^{ad}g_{ik}\delta g^{ik}) \nonumber \\
& =\int_{\partial \mathcal{M}} d^{D-1}y\epsilon\sqrt{h}f'(-\mathit{K}_{ij}\delta h^{ij}+\frac{1}{\sqrt{h}}\delta(2\mathit{K}\sqrt{h})+\mathit{K}h_{ab}\delta h^{ab}+D_iU^i)\nonumber \\
&= -\int_{\partial \mathcal{M}} d^{D-1}y f' \Pi_{ab}\delta h^{ab}+\int_{\partial \mathcal{M}}d^{D-1}y\epsilon f'\delta(2\mathit{K}\sqrt{h}) \nonumber \\
&+\int_{\partial \mathcal{M}} d^{D-1}y\epsilon\sqrt{h}D_i(f'U_i)-\int_{\partial \mathcal{M}}d^{D-1}y\epsilon\sqrt{h}(D_if')U^i
\label{a3}
\end{align}
where $ D_i $ is the spatial-covariant derivative defined on $\partial \mathcal{M}$, $ U^i\equiv n_jh^i_k\delta g^{jk} $ and for the first equality see \cite{padmanaban}. The third term of (\ref{a3}) is zero assuming the manifold is compact in D-1 dimension. The last term can be written as
\begin{align}
 -\int_{\partial \mathcal{M}} d^{D-1}y\epsilon\sqrt{h}(D_if')U^i&=-\int_{\partial \mathcal{M}} d^{D-1}y \epsilon\sqrt{h}h^e_i\nabla_ef'n_jh^i_k\delta g^{ik} \nonumber\\
&=-\int_{\partial \mathcal{M}} d^{D-1}y\epsilon\sqrt{h}\nabla_kf'n_j\delta g^{jk}
\end{align}
Then we have (\ref{a3}) as
\begin{equation}
-\int_{\partial \mathcal{M}} d^{D-1}y f' \Pi_{ab}\delta h^{ab}+\int_{\partial \mathcal{M}} d^{D-1}y\epsilon f'\delta(2\mathit{K}\sqrt{h})-\int_{\partial \mathcal{M}} d^{D-1}y\epsilon\sqrt{h}\nabla_kf'n_j\delta g^{jk}
\label{aa7}
\end{equation}
Now let's calculate the second term of surface integral in (\ref{nmn})
\begin{align}
&-\int_{\partial \mathcal{M}} d^{D-1}y\epsilon\sqrt{h}(\nabla_af')\left[-n_d\delta g^{ad}+n^a g_{ik}\delta g^{ik}\right]\nonumber\\
 &=-\int_{\partial \mathcal{M}} d^{D-1}y\epsilon\sqrt{h}\left(\nabla_af'\right)\left[-h^a_jn_i\delta g^{ij}+n^ah_{ij}\delta h^{ij}\right]\nonumber\\
&=\int_{\partial \mathcal{M}}d^{D-1}y\epsilon\sqrt{h}(\nabla_af')\left\{ h^a_jn_i-n^ah_{ij}\right\} \delta h^{ij}+\int_{\partial \mathcal{M}} d^{D-1}y\epsilon\sqrt{h}(\nabla_af')n_k\delta g^{ak}
\end{align}
where $ h^a_jn^j=0 $, $ \delta n^j=\frac{1}{2}\epsilon n^jn_kn_e\delta g^{ke}+n_kn^j_{\ell} \delta g^{k\ell}$ and also $ \delta g^{ij}=\delta h^{ij}+\epsilon n^i\delta n^j+\epsilon n^j\delta n^i$ have been used. Eventually we can write the surface integrals of (\ref{nmn}) as
\begin{equation}
\int_{\partial \mathcal{M}} d^{D-1}y\sqrt{h}\left\{-\frac{f'\Pi_{ij}}{\sqrt{h}}+\epsilon\nabla_af'\left(h^a_jn_i-n^ah_{ij}\right)\right\}\delta h^{ij}+\int_{\partial \mathcal{M}} d^{D-1}y\epsilon f'\delta(2\mathit{K}\sqrt{h})
\label{a5}
\end{equation}
Substituting (\ref{a5}) and (\ref{aa7}) in (\ref{nmn}), yields
\begin{align}
 \delta \int_{\mathcal{M}} d^Dx\sqrt{-g}f(\mathit{R})&=\int_{\mathcal{M}} d^Dx\sqrt{-g} L_{ab}\delta g^{ab}\nonumber\\
&+\int_{\partial \mathcal{M}} d^{D-1}y\sqrt{h}\left\{-\frac{f'\Pi_{ij}}{\sqrt{h}}+\epsilon\nabla_af'(h^a_jn_i-n^ah_{ij}) \right\}\delta h^{ij}\nonumber\\
&+\int_{\partial \mathcal{M}}d^{D-1}y\epsilon f'\delta(2\mathit{K}\sqrt{h})
\end{align}
where
\[ L_{ab}\equiv -\frac{1}{2}fg_{ab}+f'\mathit{R}_{ab}-\nabla_a\nabla_bf' +\square f'g_{ab}.  \]
\setcounter{equation}{0}
\section{Variation of the scalar curvature}
It is instructive to find what does the condition $ \delta\mathit{R}\mid_{\text{\tiny Boundary}}=0 $ mean. To answer this question, first let's remind the Gauss-Codazzi equation:
\begin{align}
\mathit{R}&= \ ^{(D-1)}\mathit{R}-\epsilon\{\mathit{K}_{mn}\mathit{K}^{mn}-\mathit{K}^2-2\nabla_i(\mathit{K}n^i+a^i)\} \nonumber\\
&=\ {}^{(D-1)}\!\mathit{R}+\mathit{K}^2-\mathit{K}_{mn}\mathit{K}^{mn}-2(\nabla_i\mathit{K})n^i-2n^a \nabla_i \nabla_a n^i
\label{r}
\end{align}
where ${}^{(D-1)}\mathit{R}$ is the scalar curvature of the $(D-1)$-dimensional subspace and in the first line $ a^i=n^a\nabla_an^i $ is the acceleration of the normal vector field. Then taking variation of (\ref{r}) \textbf{with $ \epsilon=-1 $} gives the Palatini identity as follows
\begin{align}
\delta\mathit{R}&={\delta\ {}^{(D-1)}}\!\mathit{R}+2\delta\mathit{K}^{mn}\left(\mathit{K}h_{mn}-\mathit{K}_{mn}\right)-2\mathit{K}\mathit{K}_{mn}\delta h^{mn}-2\nabla_i(\delta\mathit{K})n^i\nonumber \\
&-2(\nabla_i\mathit{K}+\nabla_a \nabla_i \nabla^a )\delta n^i-2n^i\delta (\nabla_a \nabla_i n^a),
\label{dr}
\end{align}
where the variation of spatial scalar curvature reads
\begin{equation}
{\delta\ {}^{(D-1)}}\mathit{R}=\ {}^{(D-1)}\!\mathit{R}_{ij}\delta h^{ij}+D_aD_d(-\delta h^{ad}+h^{ad}h_{ik}\delta h^{ik}).
\label{3r}
\end{equation}
As is obvious from (\ref{dr}) and (\ref{3r}), $ \delta\mathit{R} $ is a combination of $ \delta h^{mn},\delta n^i,\delta\mathit{K}^{mn} $ $ ,\nabla_i(\delta\mathit{K}),\delta (\nabla_a \nabla_i n^a) $ and spatial-covariant derivatives of $ \delta h^{mn} $.
\section{The conjugate momenta in $f(R)$-gravity}
%
%As we know, in order to define the momentum, by adding a scalar field $ \phi $, the Lagrangian of the $f(R)$-gravity which is not degenerate, should be made degenerate. To do so, 
To find the conjugate momenta, we write the $f(R)$ action in the Brans-Dicke form and then using the Holographic relation as in (24), make it degenerate.
 %We write the $f(R)$ action in the Branss-Dicke form and then using the Holographic relation or the Gauss–Codazzi equations, make it degenerate. This is done for the Holographic relation in (24). 
To do this, substituting the Gauss–Codazzi equation in D dimension, (\ref{r}), into (\ref{p11}), we find that 
\begin{align}
 \mathit{S}&=\int_{\mathcal{M}} d^Dx N\sqrt{h}\big\{\phi(^{(D-1)}\mathit{R}-\epsilon\{\mathit{K}_{ij}\mathit{K}^{ij}-\mathit{K}^2\})-V(\phi)\big\}\nonumber \\
&+2\int_{\mathcal{M}} d^D x\sqrt{-g}\epsilon\nabla_i(\mathit{K}n^i+a^i)\phi
\end{align}
 By-part integration on the last term gives
\begin{align}
 \mathit{S}=&\int_{\mathcal{M}} d^DxN\sqrt{h}\left\{\phi(^{(D-1)}\mathit{R}-\epsilon\{\mathit{K}_{ij}\mathit{K}^{ij}-\mathit{K}^2\})-V(\phi)\right\} \nonumber \\ 
  -&2\int_{\mathcal{M}} d^DxN\sqrt{h}\epsilon\mathit{K}n^i\nabla_i\phi-2\int_{\mathcal{M}} d^Dx\sqrt{-g}\epsilon a^i\nabla_i\phi \nonumber \\
 +&2\int_{\partial\mathcal{M}} d^{D-1}y\sqrt{h}\epsilon n_i(\mathit{K}n^i+a^i)\phi\nonumber\\=&\int_{\mathcal{M}} d^DxN\sqrt{h}\left\{\phi(^{(D-1)}\mathit{R}-\epsilon\{\mathit{K}_{ij}\mathit{K}^{ij}-\mathit{K}^2\})-V(\phi)\right\}\nonumber\\-&2\int_{\mathcal{M}} d^DxN\sqrt{h}\epsilon \mathit{K}D\phi-2\int_{\mathcal{M}} d^Dx \sqrt{-g}\epsilon a^i\nabla_i\phi+2\int_{\partial\mathcal{M}} d^{D-1}y\sqrt{h}\mathit{K}\phi \nonumber\\
\end{align}
where $ n_ia^i=0 $, $ n_in^i=\epsilon $ and $ D\phi\equiv n^i\nabla_i\phi $ have been used. Using following calculation
\begin{align}
  a_i=n^m\nabla_mn_i&=-n^m\nabla_m(N\nabla_it)=\frac{1}{N}n_in^m\nabla_mN+Nn^m\nabla_i(\frac{-1}{N}n_m)\nonumber\\
  &=\frac{1}{N}(\nabla_iN+n_in^m\nabla_mN)=\frac{1}{N}h^m_i\nabla_mN=\frac{1}{N}D_iN\nonumber\\
\label{4}
 \end{align}
 and $ n^i=(\frac{1}{N},\frac{-N^{\alpha}}{N}) $, we have 
 \begin{equation}
 ND\phi=Nn^0\partial_0\phi+Nn^{\alpha}\partial_{\alpha}\phi=\dot{\phi}-N^{\alpha}\partial_{\alpha}\phi
 \label{5}
 \end{equation}
Substituting (\ref{4}) and (\ref{5}) into (\ref{3}), we obtain
 \[\mathit{S}=\int_{\mathcal{M}} d^Dx\sqrt{h}\{N\phi(^{(D-1)}\mathit{R}-\epsilon\{\mathit{K}_{ij}\mathit{K}^{ij}-\mathit{K}^2\})-2\epsilon \mathit{K}(\dot{\phi}-N^{\alpha}\partial_{\alpha}\phi) \]
 \begin{equation}
-2\epsilon h^{ab}D_aND_b\phi-NV(\phi)\}+2\int_{\partial\mathcal{M}}d^{D-1}y\sqrt{h}\mathit{K}\phi
 \end{equation}
 Now we can define the momenta conjugate to $ h_{\alpha\beta} $, $ N $, $ N^{\alpha} $ and $ \phi $ as
\[ \bar{\Pi}_N=\bar{\Pi}_{N^{\alpha}}=0 \]
\[ \bar{\Pi}_\phi=-2\epsilon \sqrt{h}\mathit{K} \]
 \begin{equation}
 \bar{\Pi}_{ij}=\epsilon \sqrt{h}\left\{\phi(\mathit{K}_{ij}-\mathit{K}h_{ij})+\frac{h_{ij}}{N}(\dot{\phi}-N^{\alpha}\partial_\alpha\phi)\right\}
 \end{equation}
% 

%_____________________________________________________________________________________________
\section*{Acknowledgements}
F. Shojai is grateful to the University of Tehran for supporting this work under a grant provided by the university research council.
%______________________________________________________________________________________________________


\begin{thebibliography}{99}
\bibitem{dirac}{P. A. M Dirac, Phys. Rev. 114 (1959}
\bibitem{dewitt}{B. S. Dewitt, Phys. Rev. 160 (1967)}
\bibitem{york}{J. W. York, Jr, Phys. Rev. Lett. 28, 1082 (1972)}
\bibitem{reggi}{T. Regge, C. Teiltelboim, Annals Phys. 88 (1974)}
\bibitem{GHY}{G. W. Gibbons and S. W. Hawking, Phys. Rev. D 15, 2752 (1977)}
\bibitem{Hint}{E. Dyer and K. Hinterbichler, Phys. Rev. D 79:024028 (2009)}
\bibitem{fr}{A. Guarnizo, C. Leonardo, and J. M. Tejeiro, Gen.Rel.Grav. 42, 2713 (2010)}
\bibitem{myers}{R. C. Myers, Phys. Rev. D 36 (1987)}
\bibitem{madore}{N. Deruelle, J. Madore {\tt arXiv:gr-qc/0305004}}
\bibitem{merino}{N. Deruelle, N. Merino, R. Olea, {\tt arXiv:1709.06478 }}
\bibitem{krishna}{C. Krishnan, K.V.P. Kumar, A. Raju, JHEP ,10, 043 (2016)}
\bibitem{gold}{H. Goldstein, C. Poole, J. Safko, \textit{Classical Mechanics} , (2000)}
\bibitem{higher}{Hans-Jurgen Schmidt, Phys. Rev. D 49, 6354 (1994)}
\bibitem{higher1}{R. P. Woodard, Scholarpedia 10, no.8, 32243 (2015)}
\bibitem{krishnan}{C. Krishnan, and R. Avinash,  Mod. Phys. Lett. A 32, 1750077 (2017)}
\bibitem{padmanaban}{T. Padmanabhan, \textit{Gravitation: foundations and frontiers}, Cambridge University Press, (2010)}
\bibitem{padmanaban2006}{A. Mukhopadhyay and T. Padmanabhan, Phys. Rev. D 74, 124023 (2006)}
\bibitem{cap}{Salvatore Capozziello,   Valerio Faraoni, \emph{Beyond Einstein Gravity, A
Survey of Gravitational, Theories for Cosmology and Astrophysics}, Springer (2011)}
\bibitem{so}{T. P. Sotiriou, Class. Quantum Grav. 23, 17 (2006)}
\bibitem{st}{V. Faraoni, \textit{Cosmology in scalar-tensor gravity}, Springer Science and Business Media, (2004)}
\bibitem{fl}{K. Bhattacharya and B. R. Majhi, Phys. Rev. D 95, 064026 (2017)}
\bibitem{olmo}{G. J. Olmo and H. S. Alepuz, Phys. Rev. D 83, 104036, (2011)}
\bibitem{living}{A. De Felice and S. Tsujikawa, Living Rev. Rel. 13, 3 (2010)}
\bibitem{thomas}{T. P. Sotiriou, V Faraoni, Rev. Mod. Phys. 82:451-497 (2010)}
\bibitem{saltas}{I.D. Saltas, M. Hindmarsh, Class.Quant.Grav. 28, 035002 (2011)}
\bibitem{Poisson}{E. Poisson, \textit{An advanced course in general relativity}, lecture notes at University of Guelph (2002)}
\end{thebibliography}
\end{document}